# W MASS AT LEP2

P. PEREZ
DSM/Dapnia/SPP, C.E.A. Saclay, F-91191 Gif-sur-Yvette, France
On behalf of the LEP Collaborations

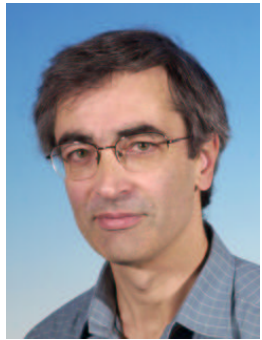

Recent studies to finalize the systematic error estimates on the measurement of the mass of the W boson at LEP2 are reviewed. Results including a new preliminary value from Aleph are updated together with the world average which is now $80.426 \pm 0.034 \text{GeV}/c^2$. The updated electroweak fit gives a 95% C.L. upper limit on the mass of the Higgs boson of $211 \text{GeV}/c^2$.

## 1  Introduction

The value for the mass of the Higgs boson is constrained from below with direct searches and from above with the measurement of observables of the standard model. The LEP1 electroweak fits give an indirect measurement of $M_{top}$ with an error of 10 GeV/$c^2$ and $M_W$ with an error of 32 MeV/$c^2$. The corresponding direct measurements at Fermilab and at LEP2 have errors of 5 GeV/$c^2$ and 34 MeV/$c^2$.

The top, W and Higgs masses are related through the following formula:
$$M_W^2 \left(1 - \frac{M_W^2}{M_Z^2}\right) = \frac{\pi \alpha}{G_F \sqrt{2}} (1 - \Delta r)$$, where $M_{top}$ and $M_H$ are hidden in the electroweak correction $\Delta r$. While $M_{top}$ and $M_H$ are related with a square root, $M_{top}$ and $M_H$ are related with a logarithm making it necessary to measure $M_W$ with such a high precision.

At LEP2, $M_W$ is measured above the pair production threshold in all its decay channels. Events are distributed among the fully hadronic or qqqq (45.6%), "semi-leptonic" or qq$\ell v$ (43.8%) and fully leptonic or $\ell v \ell v$ (10.8%) categories. For each event the measured particles 4-vectors are summed into leptons and jets according to their particle identification and event topology. The jet and lepton 4-vectors with their estimated errors enter a kinematic fit with an important constraint from the LEP beam energy. This constraint enables an improvement on the resolution from 6 or 8 GeV/c$^2$ to 2 or 3 GeV/c$^2$ on the measured W masses, depending on experiments and decay channels. For each event variables are formed, like the measured masses, their estimated errors and variables related to the topology of the event. The latter enable a rough classification for instance into four or five jet events. The distributions of these variables are compared with corresponding distributions obtained from simulations which depend on the parameter $M_W$.

In the following, all data obtained at LEP2 have been used except for OPAL which did not yet release results from year 2000 data. This represents per experiment about 700 pb$^{-1}$ of integrated luminosity, or 4500 qqqq events and 4000 qq$\ell v$ events. The LEP2 statistical errors amount to 32 and 29 MeV/c$^2$ respectively for the qq$\ell v$ and qqqq channels. The $\ell v \ell v$ channel and the method using the production cross section will not be discussed here.

## 2    Systematic errors

The main sources are errors coming from theory, beam energy, fragmentation, final state interactions (FSI) and detector biases.

*2.1) Theory:* QED corrections in the initial state modify the boost of the WW system and thus bias the result of the kinematic fit. The mass distribution itself is also affected. The difference with or without $\mathcal{O}(\alpha)$ corrections is less than 10 MeV/c$^2$. Comparisons between the RacoonWW and YFSWW [1] programs are underway. This effect is correlated between experiments and channels.

***2.2) Beam energy:*** the determination of the beam energy is reported in C. Rosenbleck's talk. The resonant depolarization technique extensively used at LEP1 is workable up to 60 GeV. Extrapolation to LEP2 physics energies is performed with magnetic field measurements (NMR), with linearity being checked with flux-loop measurements. The uncertainty is dominated by the spread between methods, $\Delta E_{beam}$=20-25 MeV. This translates into $\Delta M_W$=17 MeV/c$^2$. It is correlated between channels and considered to be correlated between experiments. Cross-checks and improvements have been achieved on the synchrotron tune and the LEP spectrometer which measures the beam energy from beam deflection and from beam energy losses.

***2.3) Fragmentation*** models have been tuned by each experiment to describe Z data after removal of b quark events. Differences between the JETSET, ARIADNE and HERWIG programs yield an estimate of 18 MeV/c$^2$ for this error which is correlated between channels and experiments. Ideas circulate on improving the understanding of this source of systematic. One possibility would be to change the rate of baryons via reweighing within ranges constrained by tunes obtained with Z data events. Another idea is to use mixed Z events and make them resemble WW events after appropriate boosts. This is also work in progress.

***2.4) FSI:*** as W's decay 0.1 fm apart on average, a scale of the order of that of strong interactions, their decay products can interact, but this effect is not built in the standard Monte Carlo programs. The mass distributions would be altered if such interactions occur between partons or particles originating from different W's, as is possible in the four quark channel. The phenomenon known as colour re-connection (CR) corresponds to the exchange of colour singlets between partons from different W's. Bose-Einstein correlations (BEC) correspond to an enhancement of identical bosons close in phase space due to the symmetrisation of their amplitude. This last effect has been observed at the Z and inside W's. If it were present also between W's, this would lead also to a distortion of the mass distribution.

Phenomenological models predict effects on the particle flow between W's. The particle population in between the jets "originating" from a given W would be relatively depleted w.r.t. the region between jets coming from different W's. This would lead to a shift in $M_W$ of 300 MeV/c² in the SKI [2] model with 100% probability for colour re-connection. A 1$\sigma$ upper limit on its parameter: $k_i$ = 2.13 is obtained from LEP data. There is no sensitivity to the

Table 1: CR models shifts

| Model | $\Delta M_W$ (MeV/c²) |
|---|---|
| SKI ($k_i$=2.13) | 74-105 |
| Herwig (CR) | 30-40 |
| Ariadne2 (CR) | 70-80 |
| Rathsman | 40-60 |

Ariadne 2 and Herwig [3] models. The shift predicted by SKI is largest and taken as an estimate for this systematic on $M_W$. It is correlated between experiments. The former number of 40 MeV/c² before summer 2002 was based on suggestions from theorists.

A possibility to constrain more the CR models is to use the high statistics Z data. OPAL and ALEPH measure the charge of gluon jets with a large rapidity gap in Z → 3 jets. L3 looks for asymmetries in particle distributions between jets. The Ariadne and Rathsman [4] models are disfavoured with a too high CR rate. Herwig does not reproduce any jet charges. The SKI model is not available for Z events. A specialized talk by M. Giunta is given on this subject in the QCD session.

Eventually, the best observable for the CR effect might be the W mass itself. $M_W$ is measured after removing either low momentum particles (pcut analysis), or particles far away from the jet axis (cone analysis), which are expected to be the most affected by the CR effect. Delphi gets $\Delta M_W$ (standard –cone) = 59 ± 35 ± 21 MeV/c² and proposes to combine such analyses on $M_W$ with Particle Flow analyses [5]. They constrain the SKI parameter in the range [0.66, 4.5]. Similar analyses are underway in the other collaborations and could be combined.

BE correlations between different W's are studied via 2-particle correlations in 4q events versus two "mixed" $\ell\nu qq$ events. L3 and ALEPH have no hint for BEC while DELPHI sees evidence for it. This inconsistency is under investigation. Although the combined result from Aleph, Delphi and L3 allows for a fraction of 0.24 ± 0.14 of the full LUBOEI (JETSET) model [6], the full shift is taken as a conservative estimate for the uncertainty on the W mass: $\Delta M_W$ = 35 MeV/c², correlated between experiments.

*2.5) Detector systematics* are determined with simulations. In practice Z peak data are used to correct Monte Carlo imperfections. The systematic uncertainties come from the statistical accuracy on a given correction and the effect of changes in the detector simulations. In the case of qq$\ell\nu$ channels, errors vary between 5 and 35 MeV/c$^2$ depending on the experiment and channel. The overall combination gives an error of 14 MeV/c$^2$. The 4q channel errors vary between 5 and 35 MeV/c$^2$ with a combined error of 10 MeV/c$^2$. These detector errors are uncorrelated between experiments.

**Table 2 : Systematics for LEP combination in MeV/c$^2$**

| | |
|---|---|
| Theory | 10 |
| LEP | 17 |
| Fragm. | 18 |
| BEC | 3 |
| CR | 8 |
| Detector | 14 |
| Total Syst. | 31 |
| Stat. | 29 |

A new preliminary analysis from Aleph [9] shows the sensitivity of the measurement of $M_W$ to the quality of the simulation. Aleph's standard simulation uses a parameterization of electromagnetic showers which gives a good description of the core of these showers. While the full EGS simulation describes better shower fluctuations and shower satellites, it is not perfect yet and cannot be improved at this stage. The collaboration decided to make the reconstruction less sensitive to this effect. An estimate on the amplitude of this effect is given by removing badly simulated clusters: $M_W$ is reduced by 100 MeV/c$^2$. A more refined event "cleaning" removes one stack ECAL clusters which represent ~2% of the total energy, and clusters around electrons and muons with an angular cut increased from 2$^o$ to 8$^o$ w.r.t. the standard analysis. This procedure is checked for the stability of the W mass measurement in the qq$\ell\nu$ channels as a function of a varying threshold on the momentum of particles. The stability is indeed much improved especially in the electron channel. Detector simulation uncertainties, after comparing full EGS with parameterisation, are increased from 20 to 30 MeV/c$^2$ in the qq$\ell\nu$ channel and from 15 to 25 MeV/c$^2$ in the 4q channel. The shift on $M_W$ after cleaning is about - 50 MeV/c$^2$ (4q), - 85 MeV/c$^2$ ($\ell\nu$qq) and -150 MeV/c$^2$ in the e$\nu$qq channel.

The systematic errors on the LEP combination are summarized in table 2.

## 3 Results

The new Aleph preliminary W mass result, for LEP energies in the interval 183 to 208 GeV, is $M_W$ (all channels)= $80.385 \pm 0.042(\text{stat}) \pm 0.041(\text{syst})$ GeV/c$^2$, $M_W(\ell\nu qq) = 80.375 \pm 0.062$ GeV/c$^2$ and $M_W$ (4q)= $80.431 \pm 0.117$ GeV/c$^2$, where the LEP prescription is used for the FSI systematic.

The overall LEP result, including threshold measurements, and the combination of LEP2 and Tevatron (Run I) results are given in table 3. The weight of the 4q channel is now 9 % (was 27 % before summer 2002) because of the increase in the FSI error. These errors might well diminish in the near future when measurements are combined allowing for stronger constraints.

**Table 3: LEP2 and World Average Results in GeV/c2**

| LEP2 | |
|---|---|
| $M_W$ (all channels) | $80.412 \pm 0.029$ (stat) $\pm 0.031$(syst) |
| $\Gamma_W$ | $2.150 \pm 0.068$ (stat) $\pm 0.060$ (syst) |
| $M_W$ ($\ell\nu qq$) | $80.411 \pm 0.032$ (stat) $\pm 0.030$ (syst) |
| $M_W$ (4q) | $80.420 \pm 0.035$ (stat) $\pm 0.101$ (syst) |
| $\Delta M_W$ (qqqq-$\ell\nu qq$) | $+22 \pm 43$ MeV/c$^2$ |
| World Average | |
| $M_W$ | $80.426 \pm 0.034$ |
| $\Gamma_W$ | $2.139 \pm 0.069$ |

If no FSI error were included, the statistical error would be 22 MeV/c$^2$. The LEP and world average values of $M_W$ are reduced by 35 and 22 MeV/c$^2$ w.r.t. summer 2002.

## 4 Interpretation

The global electroweak fit is a combination of Z-pole results (LEP + SLC), direct $M_W$ and $M_{top}$ measurements, $\nu$N scattering (NuTeV) and atomic parity violation results (with revised theory corrections). The fit has $\chi^2/\text{ndof} = 25.5/15$ (4.4% prob.). If the measurement with the largest contribution (NuTeV) is removed from the fit, one obtains $\chi^2/\text{ndof} = 16.7/14$ (27.3% prob.).

## 5  Summary

While LEP $M_W$ results are still preliminary a LEPwide effort on systematics is underway with errors from theory still preliminary and conservative, while some detector effects might have been underestimated. A new ALEPH preliminary result has been presented which changes the LEP average to

$M_W$ (LEP) = 80.412 ± 0.029 (stat) ± 0.031 (syst) GeV/c$^2$ and the world average to
$M_W$ (world) = 80.426 ± 0.034 GeV/c$^2$.

The global electroweak fit gives then an upper limit on the mass of the Higgs boson:
$M_H$ < 211 GeV/c$^2$  (95% C.L.) [8].

## References

1. A. Denner et al., Nucl. Phys. **B560** (1999) 33 ; Nucl. Phys. **B587** (2000) 67 ; Phys. Lett. **B475** (2000) 127; EPJdirect Vol.2 **C4** (2000) 1.
   S. Jadach et al., Comput.Phys.Commun. **140** (2001) 432-474.
2. T. Sjöstrand and V.A. Khoze, Z. Phys. **C62** (1994) 281;
   Phys. Rev. Lett. **72** (1994) 28.
3. L. Lönnblad, Comput.Phys.Commun **71** (1992) 15.
   G. Corcella et al., JHEP **0101** (2001) 010.
4. J. Rathsman, Physics Letters **B452** (1999) 364. A. Edin, G. Ingelman, J. Rathsman, Phys. Lett. **B366** (1996) 371; Z. Phys. **C75** (1997) 57.
5. DELPHI 2003-003-CONF-626
6. L. Lönnblad and T.Sjöstrand, Phys. Lett. **B351** (1995) 293;
   Eur. Phys. J. **C2** (1998) 165.
7. S. Heinemeyer and G. Weiglein. DCPT-02-154, Jan 2003; hep-ph/0301062.
8. CERN LEPEWWG/2003-01.
9. ALEPH 2003-005-CONF-2003-003.

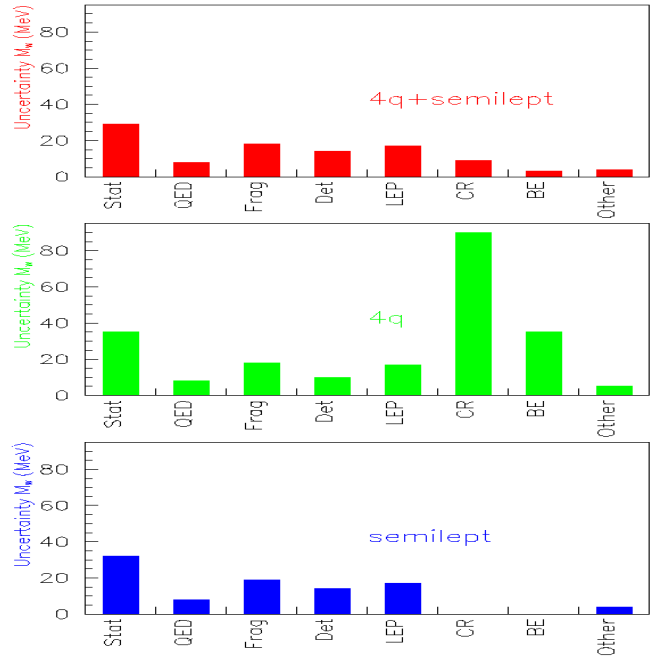

Figure 1: Summary of errors for the LEP combined result.

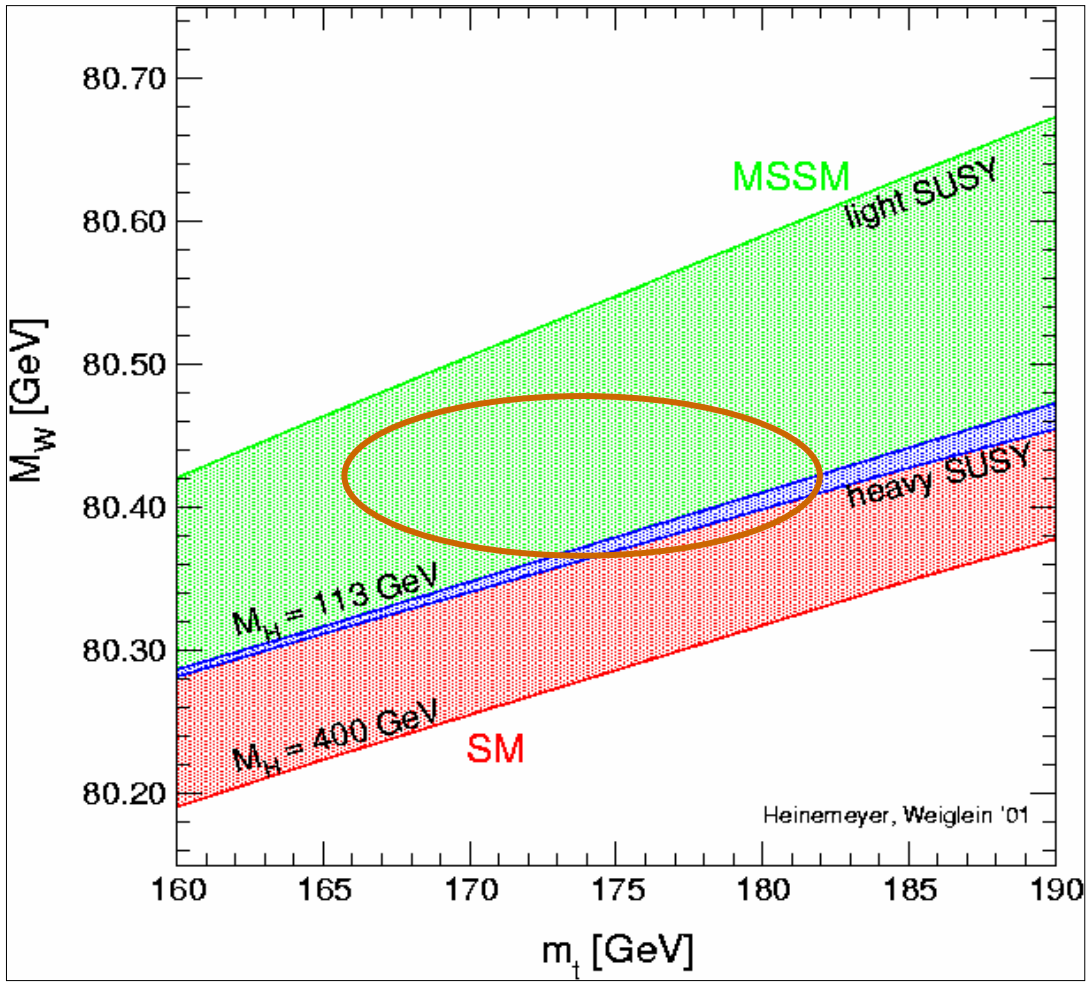

Figure 2: MW versus Mtop direct measurement with SM and MSSM predictions [7].